# Low-loss Zero-Index Materials

Haoning Tang[1,⊥], Clayton DeVault[1,⊥], Phil Camayd-Munoz[1], Yueyang Liu[2], Danchen Jia[3], Fan Du[4], Olivia Mello[1], Daryl I. Vulis[1], Yang Li[2*] and Eric Mazur[1*]

[1] School of Engineering and Applied Sciences, Harvard University, Cambridge, MA 02138, USA

[2] State Key Laboratory of Precision Measurement Technology and Instrument, Department of Precision Instrument, Tsinghua University, Beijing 100084, China

[3] Department of Optical Engineering, Zhejiang University, Hangzhou, Zhejiang, 310027, China

[4] Department of Physics, Nankai University, Nankai, Tianjin, 300071, China

[⊥] Authors contributed equally to this work

[*] Email: mazur@seas.harvard.edu

yli9003@mail.tsinghua.edu.cn

**ABSTRACT**

Materials with a zero refractive index support electromagnetic modes that exhibit stationary phase profiles. While such materials have been realized across the visible and near-infrared spectral range, radiative and dissipative optical losses have hindered their development. We reduce losses in zero-index, on-chip photonic crystals by introducing high-$Q$ resonances via resonance-trapped and symmetry-protected states. Using these approaches, we experimentally obtain quality factors of $2.6\times10^3$ and $7.8\times10^3$ at near-infrared wavelengths, corresponding to an order-of-magnitude reduction in propagation loss over previous designs. Our work presents a viable approach to fabricate zero-index on-chip nanophotonic devices with low-loss.

**INTRODUCTION**

Zero index materials have generated substantial interest in recent years because of their electromagnetic modes that exhibit stationary phase profiles.[1,2] These modes permit subwavelength confinement,[3,4] enhanced nonlinearities,[5–10] and extended quantum coherence,[11,12] opening the door to many interesting applications. Metal oxide films and metal-dielectric metamaterials exhibit near-zero index behavior at optical and near-infrared wavelength[13,14]; however, these systems have large optical losses and high impedance, which hinders the study of near-zero index phenomena and is detrimental to applications. All-dielectric photonic crystals (PhCs) eliminate metallic dissipative losses and can support a zero-index mode at $k = 0$; in addition they have finite impedance and are compatible with nanophotonic integrated circuits.[2,15,16] These properties are particularly important for on-chip nonlinear and quantum devices where a long-range, stationary phase profiles are required.[17,18]

The zero-index properties of a PhC originate from a triple mode degeneracy resulting in an accidental Dirac cone located at the gamma-point of the Brillouin zone.[2,19] Although material loss in an all-dielectric PhC is negligible, the zero-index modes are above the light-line causing substantial radiative losses both in- and out-of-plane. In-plane radiative losses can be mitigated using photonic bandgap structures, but earlier attempts[15] to reduce out-of-plane radiative losses using metallic boundaries introduce ohmic dissipation. Consequently, loss appears to be a common challenge for all zero-index materials, and novel solutions are required for zero-index phenomena and applications to come to fruition.

We substantially reduced radiative losses in near-infrared zero-index dielectric photonic crystals by increasing the quality factor ($Q$-factor) using two different approaches. The first approach achieves a high $Q$-factor by introducing resonance-trapped[20–23] modes with out-of-plane far-field destructive interference. The second approach realizes a high $Q$ via symmetry-protected modes.[24,25] Both approaches originate from bound states in the continuum;[26] however, the on-chip design breaks the out-of-plane mirror symmetry and the $Q$-factor will remain finite.[27] the devices no longer have an infinite lifetime. Characterized by a high $Q$-factor, these states have low leakage despite having an energy-momentum state which exists in the continuum of radiation. We show an improvement in quality factor and reduction in propagation loss by an order of magnitude over previous PhC Dirac cone designs. With the appropriate dielectric, our designs operate across a broad range of visible and infrared frequencies, making them versatile designs for realizing low-loss on-chip zero-index applications and devices.

**MAIN**

Our previously published, lossy zero-index PhC design consists of a square array with a 738-nm pitch of air holes of radius 222 nm in a 220-nm-thick silicon film deposited on a silicon oxide insulator substrate (Fig. 1a).[16] Using numerical finite element modeling, we determined the band structure (Fig. 1b), the $Q$-factor (Fig. 1c). The band structure near the Γ-point consists of three degenerate quasi-TE modes that have approximately linear dispersion and form an accidental Dirac cone at a wavelength of 1550 nm. At the Γ-point, the modes are classified as a degenerate pair of dipole modes and a single quadrupole mode. The $Q$-factor of the two dipole modes is $7.3 \times 10^3$ while the $Q$-factor of the quadrupole mode is $8.3 \times 10^8$, which shows that the dipole modes dominate the contribution to the structure's radiative losses.

To suppress radiative loss of the Dirac-cone modes we used two approaches to increase the $Q$-factor while retaining a Dirac-cone dispersion. The first approach involves increasing the thickness of the silicon slab to introduce a resonance-trapped dipole mode. The on-resonance condition for this design corresponds to total destructive interference between the doubly degenerate dipole modes and additional on-Γ modes radiating from the photonic crystal slab.[20–22,28,29] We optimized (see Supplementary Material Fig. S2 and Fig. S3) the airhole radius, array pitch, and slab thickness to obtain both a Dirac cone and a high-$Q$ resonance at 1550 nm, yielding a geometry of $r = 197$ nm; $p = 630$ nm; $t = 570$ nm (Fig. 1d). The geometry yields a degenerate Dirac cone dispersion at 1550 nm (Fig. 1e) with a dipole mode $Q$-factor of $9.0 \times 10^4$, an order of magnitude larger than that in the lossy design (Fig. 1f).

The second approach involves symmetry-protected modes,[30] obtained in a hexagonal lattice of flower-shaped air-holes represented by an airhole radius $r(\phi) = r_o + r_d cos(6\phi)$, with $r_o = 226.4$ nm and $r_d = 109$ nm, a unit-cell pitch of $p = 740$ nm, and a slab thickness of $t = 370$ nm (Fig. 1g). The hexagonal lattice belongs to the $C_{6v}$ point group and supports an irreducible representation of quasi-TM modes which do not couple to plane waves at normal incidence because their symmetry is incompatible with plane waves at the Γ-point.[31] Figures 1h–i show the calculated dispersion and $Q$-factors for the three modes in this symmetry-protected PhC near the Γ-point. The modes form a Dirac cone dispersion with minimum $Q$-factor of 6.0×10$^6$, which is three orders of magnitude larger than the lossy design.

Figure 2 shows the in-plane and out-of-plane (insets) field profiles of the three zero-index designs. For each design, the photonic crystal's interfaces are connected to SU-8 ridge waveguides. We numerically compute the field profiles by exciting the zero-index mode of the three structures with in-phase dipole sources located in each unit cell. All three structures support zero-index modes leading to highly directional in-plane emission into the SU-8 waveguides, facilitated by the non-zero group velocity and finite impedance and finite impedance of Dirac cone structures (see Supplementary Material Fig. S5 and Fig. S6). As expected from the $Q$-factor in Fig. 1c, the lossy structure shows significant leakage of radiation in the out-of-plane direction (Fig. 2a inset). In the resonance-trapped PhC design, radiative modes are better suppressed (Fig. 2b). Even stronger suppression is observed in the symmetry-protected PhC design (Fig. 2c). The small amount of leakage radiation visible in Figs. 2b and 2c is due to direct radiation from the dipole sources.

To verify our numerical findings, we fabricated the lossy, resonance-trapped PhC, and symmetry-protected PhC designs in a Silicon-on-substrate (SOI) wafer with a 220-nm-thick silicon device layer. To obtain the thicknesses required for the resonance-trapped (570 nm) and symmetry-protected (370 nm) designs, we used chemical vapor deposition to deposit amorphous silicon on top of the device layer; the index contrast between the crystalline silicon and deposited amorphous silicon is small (≈ 0.01).[32,33] The airhole patterns were then defined using conventional electron-beam lithography methods. Scanning electron images of the resonance-trapped and symmetry-protected PhCs are shown in Fig. 3a and 3b, respectively.

To determine the zero-index wavelength of our PhC devices, we used the Fourier microscope setup shown in Figure 3c to image the isofrequency contours[34] of our three devices (lossy, resonance-trapped PhC, and symmetry-protected PhC designs). Each device is illuminated with a polarized and collimated tunable (1500–1630 nm) laser beam using a near-infrared 10X objective. For a given wavelength, we adjust the incidence angle to match the momentum of a particular resonant mode by moving L1. Light from this resonant mode is then scattered by fabrication disorder into modes with similar momentum, which then radiate to form isofrequency contours in the far-field.[34] We remove the incident laser beam using a second polarizer (Supplementary Material Fig. S7) and then image the contours onto a CCD camera.

Figure 4 shows the measured and calculated isofrequency contours of the three samples. The color scale represents the numerically calculated $Q$-factor of each contour. For clarity, we normalize the intensity of the CCD contour images because the scattered intensity approaches zero for high-$Q$ contours. The lossy PhC design supports quasi-TE resonances with low-$Q$ dipole modes below the Dirac point wavelength of 1549.0 nm. In the experimental images (Fig. 4b), we see bright contours below 1549.0 nm, a bright large-area contour corresponding to the flat band at 1548.5 nm, and dark contours above 1549.0 nm. The resonance-trapped PhC design also supports quasi-TE resonances; however, near the Dirac point wavelength of 1548.0 nm we observe high $Q$-factors and dim scattering contours (Fig. 4c & 4d). For the symmetry-protected PhC, the resonances are quasi-TM and the $Q$-factor peaks sharply near the Dirac point wavelength of 1558.0 nm (Fig. 4e). The experimental images (Fig. 4f) show darker contours as the wavelength increases from 1530.0 to 1555.0 nm, indicating progressively smaller radiative losses. Within experimental resolution, we do not observe a bandgap in our devices. At the Dirac point wavelength of 1558.0 nm, we observe minimal scattered intensity at the center of the Fourier plane, indicating the presence of a low-loss mode at the Γ-point. The off-centered bright scattered intensity corresponds to a lower-$Q$ flat band. For wavelengths greater than 1558.0, the contour becomes bright again in good agreement with our numerical calculations.

To measure the $Q$-factor of the devices, we placed a pinhole in the Fourier plane following lens L3 in Fig. 3c. The diameter of the pinhole is 200 μm, yielding a momentum resolution of $\delta k \sim 0.002$. The pinhole is mounted on a 2-axis stage to select specific $k$-points of the iso-frequency contours. We direct the light passing through the pinhole to a photodiode. The photodiode is connected to a lock-in amplifier which is synchronized with a 1-kHz chopper placed in the incident laser beam. We then record the scattered light intensity from the lock-in amplifier, sweeping the laser wavelength and varying the pinhole position to obtain the $Q$-factor at various $k$-values. The normalized intensity measurements for the lossy, resonance-trapped, and symmetry-protected PhC designs are shown in Fig. 5a, 5c, and 5e, respectively, and the data collected at a particular $k$-point is fitted with a Lorentzian. In Fig. 5b, 5d, and 5f, the total $Q$-factors on the left vertical axis correspond to the full-width at half-maximum of the Lorentzians. The right vertical axis shows the absolute value of the index obtained from the measured contour images (Fig. 5b, d, f) using the relationship $n = ck/\omega$. Our index retrieval simulations indicate negative indices at larger wavelengths (see Supplementary Materials Fig. S5). For the lossy PhC design (Fig. 5a, 5b), we find a $Q$-factor of 351.5 near the zero-index wavelength of 1549.0 nm. At longer wavelength, the $Q$-factor substantially increases due to a higher-$Q$ off-Γ mode (Fig. 1b). The resonance-trapped PhC (Fig. 5c, 5d) has a $Q$-factor of $2.6 \times 10^3$ near a zero-index wavelength of 1548.0 nm, while the symmetry-protected PhC (Fig. 5e, 5f) has a $Q$-factor of $7.8 \times 10^3$ near a zero-index wavelength of 1558.0 nm. The results for the resonance-trapped and symmetry-protected PhC design are nearly an order-of-magnitude improvement over the lossy PhC design.

The measured $Q$-factors are lower than the calculated ones because of the finite size of the device and fabrication disorder (see Supplementary Material Fig. S8). The finite-size of the patterned area breaks the periodicity of the crystal and Bloch's theorem is no longer valid in a single unit-cell. However, if we treat the device as periodic over the length scale of the device and assume a random distribution of disorder, then we can expand the device's super-cell mode over the Bloch modes of an ideal photonic crystal evaluated at the fractional orders of the wavevector.[35,36] For the zero-index mode at the Γ-point, the super-cell mode of our photonic crystal designs contain off-Γ, fractional modes with finite $Q$-factors, and the total $Q$-factor is reduced. Our experimental results could therefore be improved by increasing the device area and reducing fabrication disorder. Radiation leakage into the SOI substrate also decreases the measured $Q$-factors. This loss could be mitigated by using an index-matching layer or suspending the devices to create a symmetric index profile in the out-of-plane mirror direction (see Supplementary Material Fig. S9).

In conclusion, we demonstrated two approaches to mitigate radiative losses in near-infrared zero-index photonic crystals. We experimentally demonstrated an order of magnitude increase in the total quality factor over previous designs, and therefore a ten-fold reduction in loss. Losses could be further reduced by improving fabrication and the index profile. The low-loss zero-index photonic crystal designs we present support on-chip operation, impedance matching, and scalability across the visible and infrared spectrum, given the appropriate high-index dielectric.[37] These attributes are beneficial for zero-index applications which require low loss and long propagation length scales (see Supplementary Material Fig. S10), such as enhanced nonlinear and quantum interactions in photonic integrated circuits. Our findings open novel routes for the development of on-chip zero-index devices and applications which have so far remained elusive due to the large losses in current designs.

**MATERIALS AND METHODS**

**Simulation**

The band structures, $Q$-factor, and isofrequency contours were computed using three-dimensional finite element method simulations (COMSOL Multiphysics 5.4). We first calculated all the modes in a PhC unit cell with Floquet periodic boundary conditions in the two lattice-vector directions and perfectly matched layers at the boundaries in the out-of-plane direction. TM/TE-polarized modes were selected by evaluating the energy ratio of the electric and magnetic fields in all directions. The field profile, effective index and impedance, propagation loss, and Fano resonance were computed using 3D finite-difference time-domain simulator (Lumerical FDTD).

**Sample Fabrication**

All devices were fabricated on 220-nm SOI wafers. Using low growth rate PECVD we deposited additional amorphous silicon to obtain the required thicknesses for the resonance-trapped PhC (570 nm) and symmetry-protected PhC devices (370 nm). To remove the oxide and enhance Si quality, we treated the wafer using RCA cleaning processes. The wafer was then coated with the positive photoresist (ZEP520A) for e-beam writing, followed by reactive ion-beam etching. The positive photoresist was removed afterwards.

## ACKNOWLEDGMENTS

The authors thank Yi Yang, Chao Peng, Renjing Xu, Orad Reshef, Carlos Rio, Meng Xiao, Shahin Firuzi, and Tian Dong for discussions. P.M. provided the idea of resonance-trapped PhC for this work. H.T., Y.Y.L., Y.L., C.D., D.F. and O.M. carried out the simulations. D.V. provided the initial design of resonance-trapped ZIM and performed initial calculations and experiments. H.T. designed and fabricated the samples. H.T, C.D. and D.J. carried out the measurements. H.T and D.J. analyzed the experimental results. Y.L. and E.M. supervised the research and the development of the manuscript. H.T. and C.D. wrote the manuscript with input from all authors. All authors subsequently took part in the revision process and approved the final copy of the manuscript. The Harvard University team acknowledges support from DARPA under contract URFAO: GR510802 for the optical characterization, numerical simulations, device fabrication described in this paper. The sample fabrication was performed at Harvard University's Center for Nanoscale Systems, which is a member of the National Nanotechnology Coordinated Infrastructure Network and is supported by the National Science Foundation under NSF award 1541959.

**FIGURES**

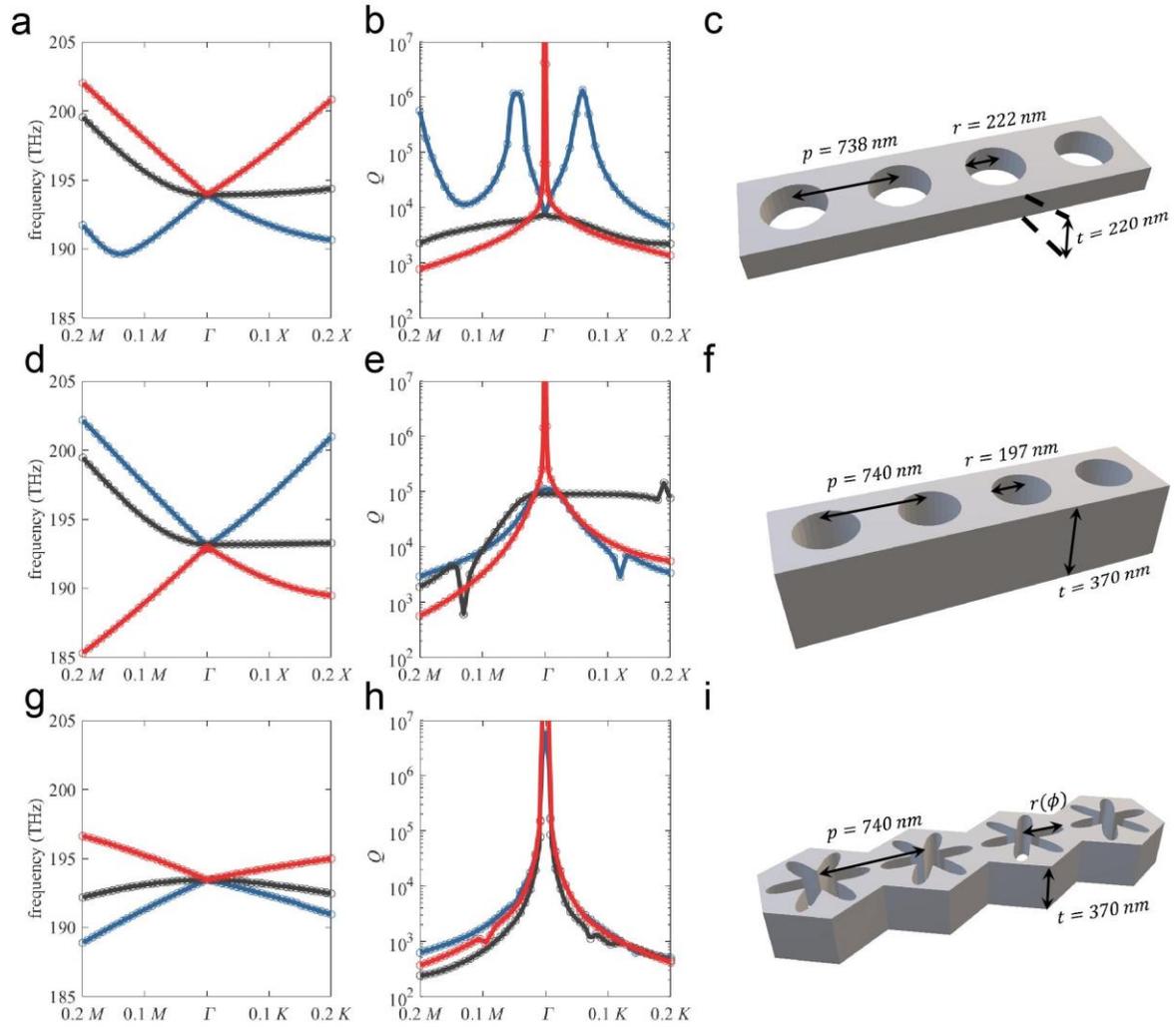

**Fig. 1    Reducing radiation losses using resonance-trapped and symmetry-protected designs.** Band structure (left), quality factor (middle) and device geometry (right) for lossy (top), resonance-trapped (middle) and symmetry-protected (bottom) zero-index PhC designs (substrates are not shown for clarity). The colors in the band structure and quality factor plots correspond to distinct modes. For the lossy and resonance-trapped PhC designs, the red curves correspond to quadrupole modes and the blue ones to dipole modes. For the symmetry-protected PhC design, the modes are more complex (see Supplementary Material Fig. S1).

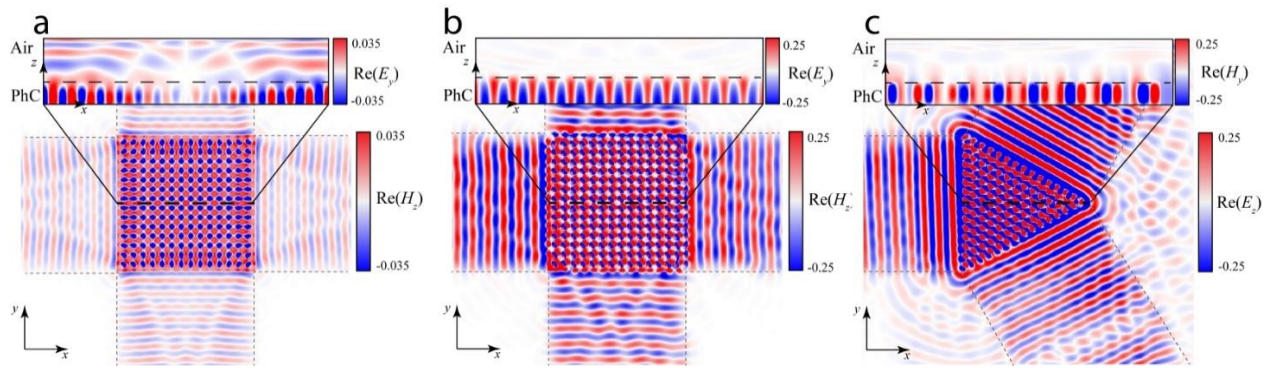

**Fig. 2   Electromagnetic field patterns in zero-index materials.** In-plane field patterns and radiative field patterns (inset) in lossy (left), resonance-trapped (middle) and symmetry-protected (right) zero-index PhC designs excited by dipole sources. In the in-plane direction, the waves are emitted from the PhC slabs into SU-8 waveguides; in the out-of-plane direction, the waves are radiated into air.

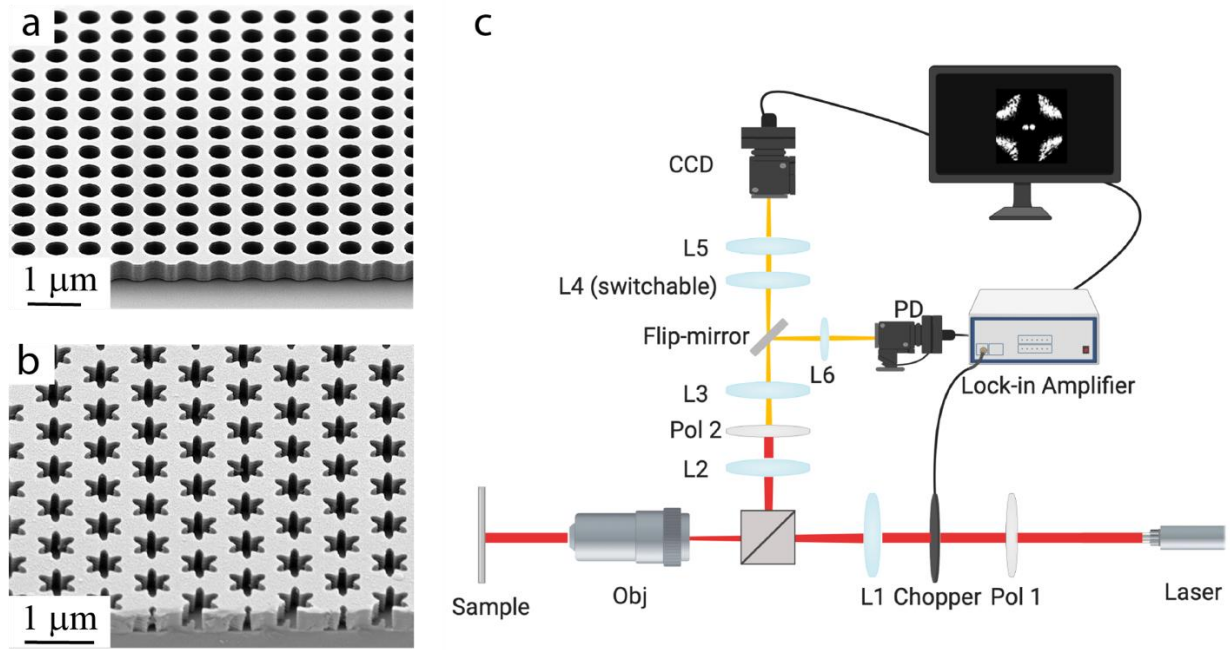

**Fig. 3    Device and experimental setup.** Scanning electron microscopy images of the (a) resonance-trapped and (b) symmetry-protected PhC devices. The total patterned area is approximately 500×500 µm². (c) Setup for measuring iso-frequency contours. A collimated laser is first polarized and then focused by lens (L1) with focal length $f = +15$ cm (L1) onto the back focal plane of a 10x infinity-corrected NIR objective, so the sample is illuminated by a collimated beam. The angle of incidence is controlled by moving L1 in the horizontal plane to achieve resonance coupling at each excitation wavelength. We image the back focal plane of the objective onto the CCD camera using a 1.67-magnifcation $4f$-relay lens system. Lens L4 can be removed to switch the setup to real-space imaging for initial alignment.

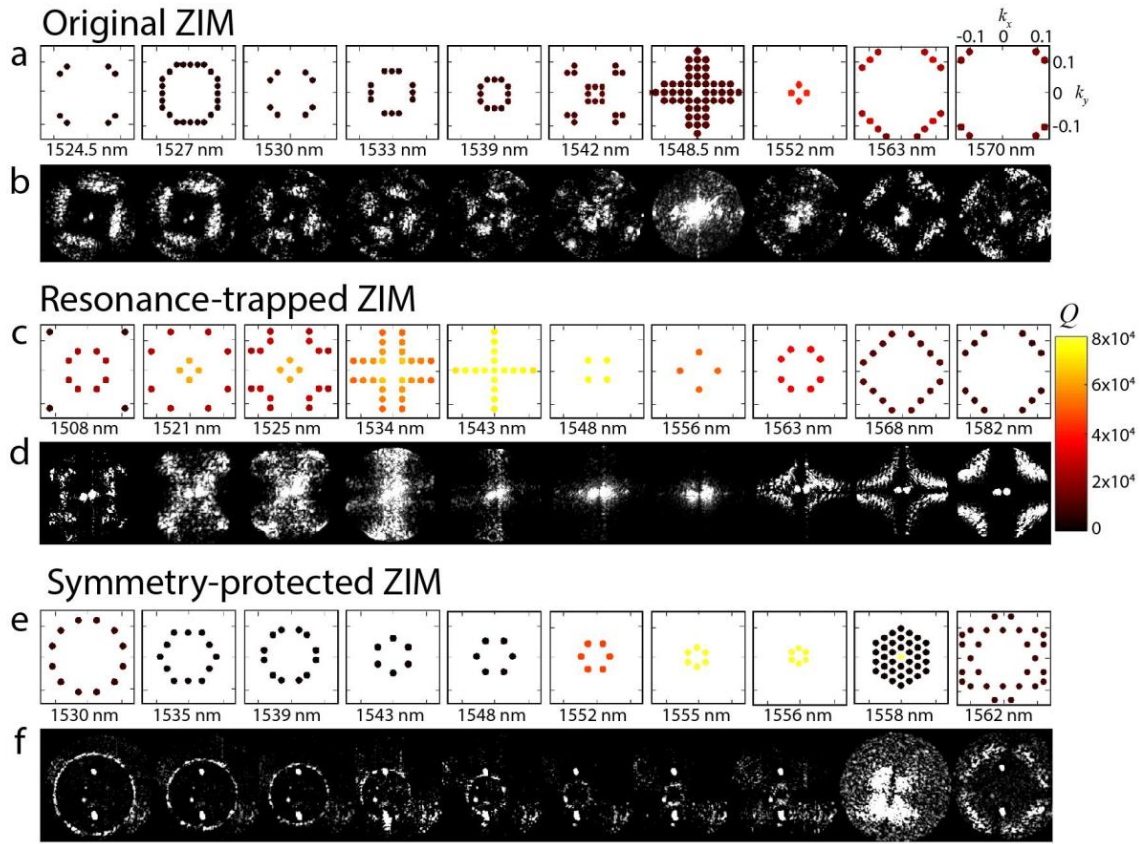

**Fig. 4   Contour images.** Numerical (a, c, e) and measured (b, d, f) iso-frequency contours of lossy (top), resonance-trapped (middle), and symmetry-protected (bottom) zero-index PhC devices. The color scale indicates the calculated $Q$-factor. In the measured images the contrast is normalized to improve visibility. The pair of bright spots visible in many of the measured images are the laser's incident and reflected beam.

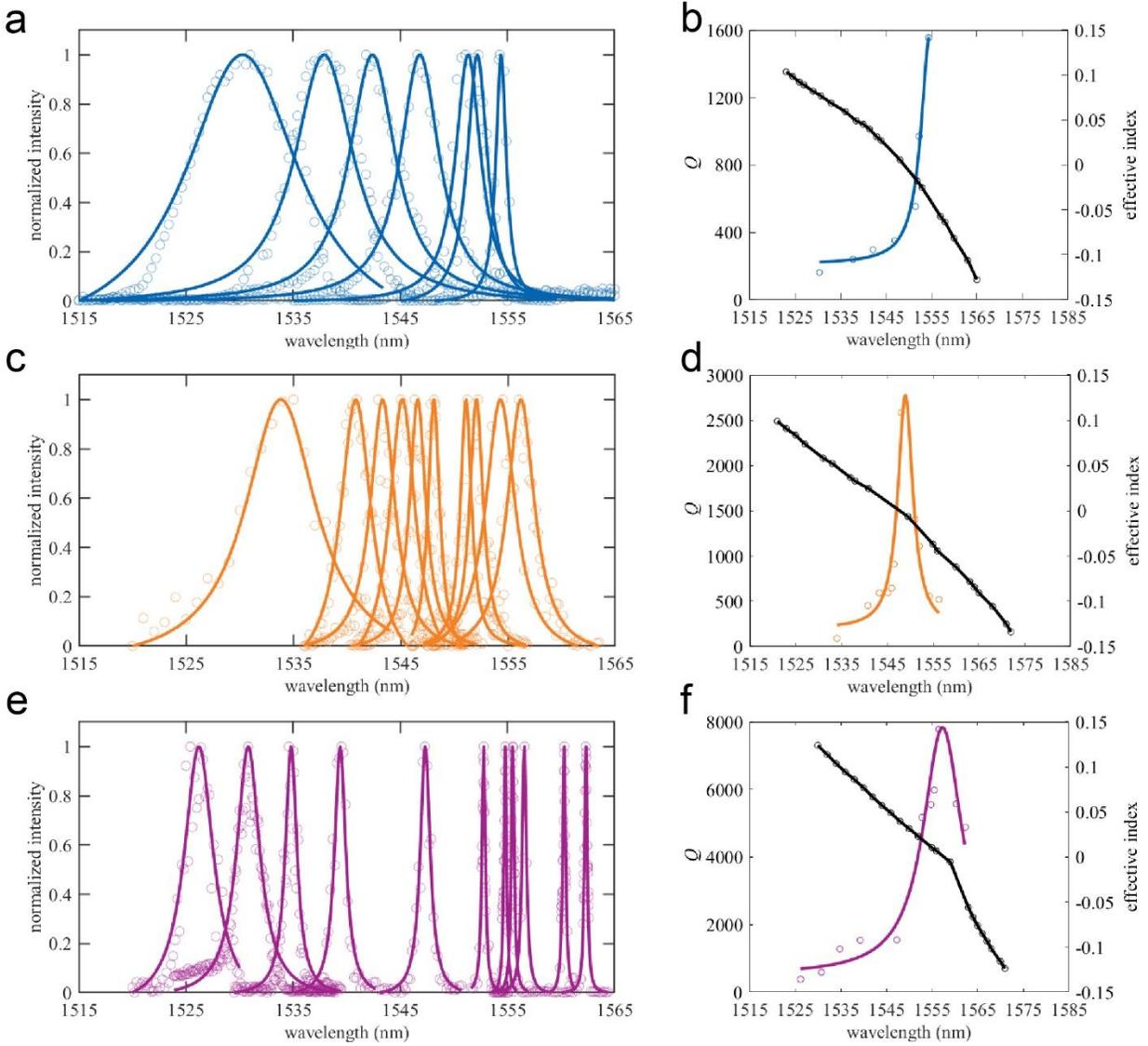

**Fig. 5   Quality factor measurement.** Lorentzian fits to the wavelength-dependence of the normalized scattered-light intensity (a, c, e); each curve represents a different pinhole position and therefore a different k-vector. The $Q$-factors (left vertical axis of b, d, f) are obtained from the Lorentzian line widths on the left. The effective indices (right vertical axis of b, d, f) are determined from the measured iso-frequency contour. (a–b) Lossy, (c–d) resonance-trapped, (e–f), and symmetry-protected PhC.

Supplementary Materials for

# Low-loss Zero-Index Materials

Haoning Tang[1,⊥], Clayton DeVault[1,⊥], Phil Camayd-Munoz[1], Yueyang Liu[2], Danchen Jia[3], Fan Du[4], Olivia Mello[1], Daryl I. Vulis[1], Yang Li[2*] and Eric Mazur[1*]

[1] School of Engineering and Applied Sciences, Harvard University, Cambridge, MA 02138, USA

[2] State Key Laboratory of Precision Measurement Technology and Instrument, Department of Precision Instrument, Tsinghua University, Beijing 100084, China

[3] Department of Optical Engineering Zhejiang University, Hangzhou, Zhejiang, 310027, China

[4] Department of Physics, Nankai University, Nankai, Tianjin, 300071, China

[⊥] Authors contributed equally to this work

[*] Email: mazur@seas.harvard.edu

  yli9003@mail.tsinghua.edu.cn

**Section 1. Create a Dirac-cone triply degeneracy**

The accidental degeneracy at the $\Gamma$-point of a Dirac-cone zero-index material (ZIM) photonic crystal (PhC) slab requires a triple degeneracy between a single mode and two degenerate modes. The lossy and resonance-trapped ZIM PhC have a $C_{4v}$ symmetry. The Dirac-cone consists of a single mode with a $B_1$ representation and two double degenerate modes with $E$ representations (Fig. S1(a-b)). All three modes are quasi-TE modes. The symmetry-protected ZIM PhC has a $C_{6v}$ symmetry. The Dirac-cone consists of a single mode with $B_1/B_2$ representation and a pair of double degenerate modes with $E_2$ representations (Fig. S1(c)). All three modes are quasi-TM modes.[1]

**Section 2. Tunable resonance-trapped ZIM PhC**

We vary the height and radius of the ZIM PhC slab to achieve a resonance-trapped mode at the Γ-point. Fig. S2 shows the $Q$-factor (color map) over the radius and height parameter space with the Dirac-cone degeneracy condition indicated by a white line. The maximum $Q$-factor along this line corresponds to the optimal height and radius of $h = 570$ nm and $r = 197$ nm, respectively. Fig. S3 shows the numerically calculated (FDTD) normal-incident transmission spectrum of our ZIM PhC slab for various height values. The spectrum consists of a Fano line shape over Fabry-Pérot background oscillations. The Fano line shape arises from the far-field interference between the radiative dipole resonance and the non-resonant PhC slab. At $h = 570$ nm the magnitude of the Fano line shape vanishes which indicates a strong suppression in the dipole mode's radiative decay.

**Section 3. Effective index, effective impedance and group index**

Our ZIM designs achieve an effective refractive index of zero along with a finite impedance and group velocity, allowing for efficient coupling between the device and standard optical waveguides. The effective index of our devices is calculated using both simulated in-plane reflection/transmission monitors and dispersion curves. In the first approach, we record the in-plane transmission and reflection spectrum using Si waveguides coupled to the PhC device boundaries (Fig. S4(a)). The fundamental mode of the Si waveguide is used to excite the zero-index mode and we retrieved the effective index and impedance of our three devices using an FDTD retrieval algorithm (Fig. S5).[2] The impedances of lossy, resonance-trapped, and symmetry-protected ZIM PhCs are 1.0, 1.0, and 0.18, respectively, at the zero-index wavelength of 1550 nm. In addition, we calculate the effective index from each of the three device's band structures using $n_e = ck/\omega_k$ (Fig. 1 and S6(a)). In addition, we calculate the group index using $n_g = v_g/c$ where $v_g$ is the group velocity (Fig. S6 (b)). The group indices at the zero-index wavelength of lossy, resonance-trapped, and symmetry-protected ZIM PhC are 7.2, 7.5, and 10.4, respectively. The finite values of impedance and group index indicate a good coupling to free space and to standard optical waveguides, showing the significance for our designs for integrated photonics circuits.

**Section. 4 Temporal Coupled Mode Theory: Line shape of the quality factor**

In our measurements, we use two orthogonal polarizers to filter the direct reflection. We then record the amplitude of light scattered into the orthogonal polarization, which is first normalized and then fitted to a Lorenz line shape to extract the total $Q$-factor. To elucidate the origin of the Lorentz line shape, we use temporal coupled mode theory.[3] Let $|a\rangle$ be the amplitude of a resonance at frequency $\omega_o$ with decay constant $\gamma$. The dynamics of the system can be written as

$$\frac{d}{dt}|a\rangle = -(i\omega_o + \gamma) + K^T|s_+\rangle \tag{S1}$$

$$|s_-\rangle = C|s_+\rangle + D|a\rangle \tag{S2}$$

The incoming ($|s_+\rangle$) and outgoing ($|s_-\rangle$) waves consist of $s$ and $p$ polarizations; we denote waves above and below the photonic crystal interface as $s_1$ and $s_2$, respectively:

$$|s_+\rangle = \begin{pmatrix} s_{+,1}^s & s_{+,2}^s & s_{+,1}^p & s_{+,2}^p \end{pmatrix}^T \tag{S3}$$

$$|s_-\rangle = \begin{pmatrix} s_{-,1}^s & s_{-,2}^s & s_{-,1}^p & s_{-,2}^p \end{pmatrix}^T.$$

(S4)

Incoming waves can couple to the resonance mode via the coupling vector $K$ given by,

$$K^T = \begin{pmatrix} k_1^s & k_2^s & k_1^p & k_2^p \end{pmatrix}; \tag{S5}$$

in the complimentary process, the resonance mode decays into outgoing waves via the coupling vector $D$,

$$D^T = \begin{pmatrix} d_1^s & d_2^s & d_1^p & d_2^p \end{pmatrix}. \tag{S6}$$

Incoming and outgoing waves can couple through a direct process which takes a block-diagonal form when operating at near-normal incidence

$$C = \begin{pmatrix} C_s & 0 \\ 0 & C_p \end{pmatrix}; \qquad C_{s,p} = \begin{pmatrix} r_{s,p} & t'_{s,p} \\ t_{s,p} & r'_{s,p} \end{pmatrix}.$$

(S7)

Assuming a time-harmonic resonance mode, Eqs. S1 and S2 become

$$|s_-\rangle = \left[ C + \frac{DK^T}{-i(\omega-\omega_o)+\gamma} \right] |s_+\rangle,$$

(S8)

where

$$DK^T = \begin{pmatrix} d_1^s \\ d_2^s \\ d_1^p \\ d_2^p \end{pmatrix} \begin{pmatrix} k_1^s & k_2^s & k_1^p & k_2^p \end{pmatrix} = \begin{pmatrix} d_1^s k_1^s & d_1^s k_2^s & d_1^s k_1^p & d_1^s k_2^p \\ d_2^s k_1^s & d_2^s k_2^s & d_2^s k_1^p & d_2^s k_2^p \\ d_1^p k_1^s & d_1^p k_2^s & d_1^p k_1^p & d_1^p k_2^p \\ d_2^p k_1^s & d_2^p k_2^s & d_2^p k_1^p & d_2^p k_2^p \end{pmatrix}.$$

(S9)

We set the incident light to linear $s$-polarization using the first polarizer, such that $|s_+\rangle = (s_{+,1}^s \ 0 \ 0 \ 0)^T$. Inserting $|s_+\rangle$ into Eq. S8, we find the outgoing wave amplitude

$$|s_-\rangle = \begin{pmatrix} s_{-,1}^s \\ s_{-,2}^s \\ s_{-,1}^p \\ s_{-,2}^p \end{pmatrix} = \left[ \begin{pmatrix} r_s \\ t_s \\ 0 \\ 0 \end{pmatrix} + \frac{1}{-i(\omega-\omega_o)+\gamma} \begin{pmatrix} d_1^s k_1^s \\ d_2^s k_1^s \\ d_1^p k_1^s \\ d_2^p k_1^s \end{pmatrix} \right] s_{+,1}^s$$

(S10)

With the second polarizer (analyzer), we select reflected linear $p$-polarized light which is the $s_{-,1}^p$ component of $|s_-\rangle$,

$$s_{-,1}^p = \frac{d_1^p k_1^s}{-i(\omega - \omega_o) + \gamma} s_{+,1}^s. \qquad (S11)$$

Eq. S11 takes the form of a Lorenz function. The measured signal is first normalized and then fitted to Eq. S11 to extract the $Q$-factor where $Q_{tot} = \frac{\tau \omega_o}{2} = \frac{\omega_o}{2\gamma}$.

**Section 5. Disorder characterization**

Fabrication disorder decreases the $Q$-factor of our devices.[4] The deep reactive ion etching process creates roughness on the structure sidewall. To analyze the degradation of the symmetry-protected device's $Q$-factor we introduced a triangular sidewall disorder in numerical simulations. The degree of disorder is characterized by a ratio of triangular height ($d$) to unit cell pitch ($R$). Fig. S8 shows the $Q$-factor (the smallest $Q$-factor among three modes) drops from $10^8$ to $10^4$ as the fabrication disorder ($d/R$) increases from 0.03 to 0.07, while maintaining the accidental degeneracy of three Dirac-cone modes.

**Section 6. Bound states in the continuum (BIC) property in PhC slab with out-of-plane mirror symmetry**

The high $Q$-factors of our devices originate from a bound state in the continuum; however, the SOI substrate required in on-chip designs introduces an asymmetry in the out-of-plane refractive index profile and reduces the total $Q$-factor of both resonance-trapped and symmetry-protected BICs.[5] The performance of our devices could be substantially improved by suspending the structures either in the air or using index matching oil to achieve the out-of-plane mirror symmetry. Fig. S9 shows the band structure and $Q$-factors of the resonance-trapped and symmetry-protected ZIM PhCs when suspended in the air. The modes are true BICs in this condition and we numerically find the $Q$-factor diverges to infinity in both resonance-trapped and symmetry-protected ZIM designs.

**Section 7. Propagation Loss**

We estimate the in-plane propagation losses of the ZIM PhC slab using the cut-back method.[6] In this simulation, we couple into and out of a PhC slab by a pair of silicon slab waveguides at the same height (Fig. S4(a)). The optical power propagating within the PhC slab was monitored at different propagation lengths, and the propagation loss was calculated from the slope of transmission versus propagation length curve. Fig. S10, the light propagates over 40 $\mu m$ within the PhC structure. The in-plane propagation loss of lossy ZIM PhC in $\Gamma$-$X$ direction is 5.6 dB/mm at the zero-index wavelength. For the resonance-trapped ZIM PhC design, the propagation loss at the zero-index wavelength is 0.83 dB/mm in the $\Gamma$-$X$ direction; for the symmetry-protected ZIM PhC it is 0.23 dB/mm in the $\Gamma$-$M$ direction. In the resonance-trapped and symmetry-protected ZIM PhCs, the small oscillations in the propagation loss are caused by reflections at the boundaries and enhanced field build up inside the PhC.[7]

**Section 8. Angle-resolved transmission/reflection spectra of out-of-plane incidence on PhC slabs**

To obtain the angle-resolved transmission/reflection spectra, we simulated a plane wave incident from the top of the PhC slab for s and p polarizations. Transmission and reflection are monitored at the bottom and top of the slab (Fig. S4(b)). For all three types of ZIM PhC, the Fano line shapes at different incident angles (Fig. S11) are in good agreement with the Dirac-cone band structure (left column). S-polarized light excites the upper and lower bands of the Dirac-cone dispersion (middle column) while p-polarized light excites the middle band of the Dirac-cone degeneracy (right column). The width of a Fano resonance at a certain combination of wavelength and incident angle is inversely proportional to the $Q$-factor of the mode at the corresponding wavelength and wavenumber. The modulation in the transmission and reflection curves is caused by a Fabry–Perot resonance in the PhC slab. Because of the low $Q$-factors for the lossy ZIM PhC design, the Fano resonance peaks have the same amplitude at all incident angles (Fig. S11(a-c)). The Fano resonance peaks vanish as the incident angle approaches zero for both resonance-trapped (Fig. S11 (d-f)) and symmetry-protected (Fig. S11(g-i)) PhCs because the modes do not interfere in the far-field as a result of the high $Q$-factor. Thus, the vanishing of the Fano line shape at 0° agrees with a high $Q$-factor at the Γ point.

**Figures**

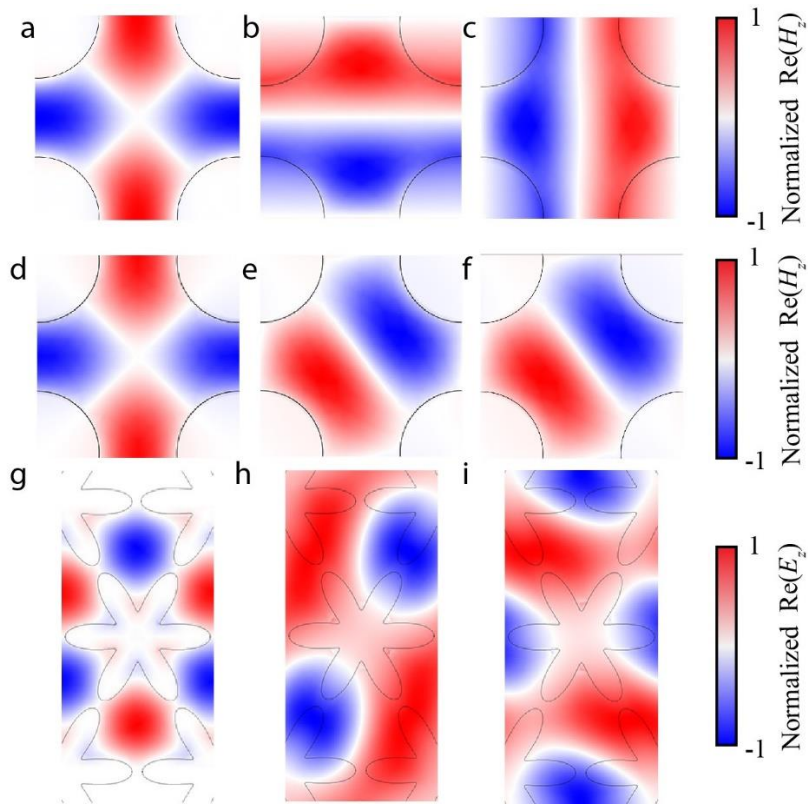

**Fig. S1 Modes in Dirac-cone triply degeneracy.** Dirac-cone triple degeneracy with one single degenerate mode (left) and two double degenerate modes (middle and right). (a-c) Magnetic field $Re(H_z)$ in the unit cell of the lossy ZIM PhC for the quasi-TE modes. (d-f) Magnetic field $Re(H_z)$ in the unit cell of the resonance-trapped ZIM PhC for the quasi-TE modes. (g-i) Electric field $Re(E_z)$ in the unit cell of the symmetry-protected ZIM PhC for the quasi-TM modes.

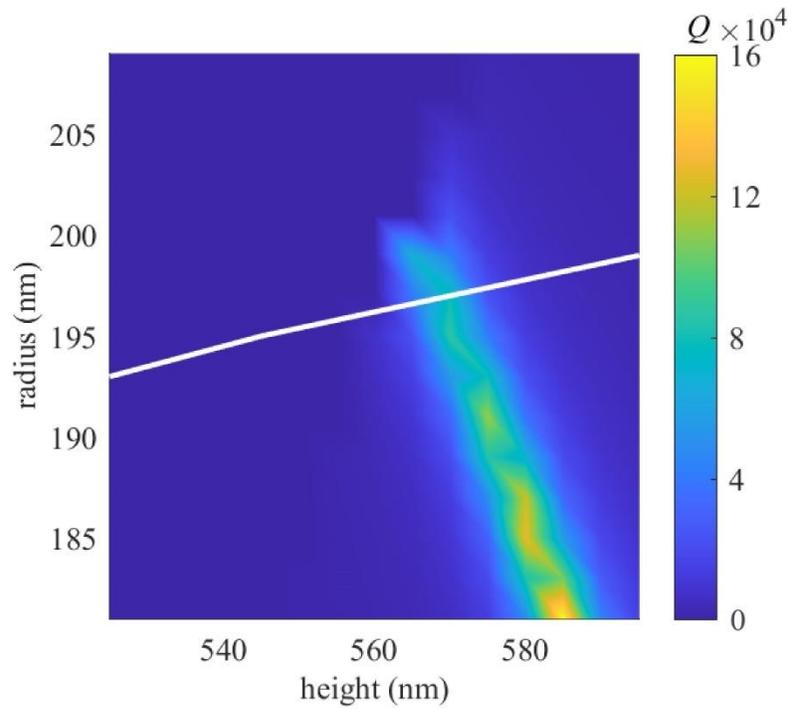

**Fig. S2 Resonance-trapped mode.** The quality factor of the dipole mode as a function of the radius and height of the ZIM PhC (Fig. 1(f)). The white line indicates the triple degeneracy of the dipole modes (Fig. S1(e, f)) and quadruple mode (Fig. S1(d)). This line crosses the high-$Q$ region at $r$ = 198 nm and $h$ = 570 nm, indicating a resonance-trapped ZIM.

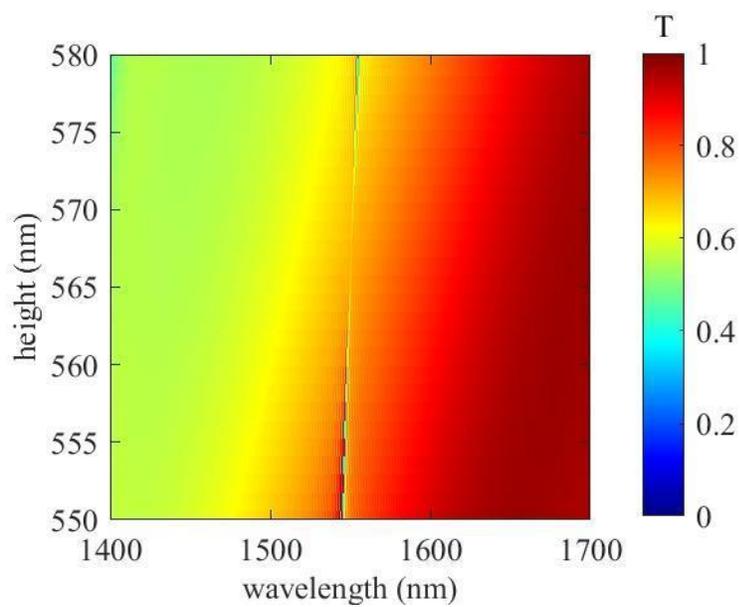

**Fig. S3 Vanishing linewidth in transmission spectra.** FDTD simulation of out-of-plane transmission spectra at normal incidence for different height of ZIM PhC (Fig. 1(f)). This result shows the vanishing linewidth of the dipole mode at a height of 570 nm because of resonance trapping.

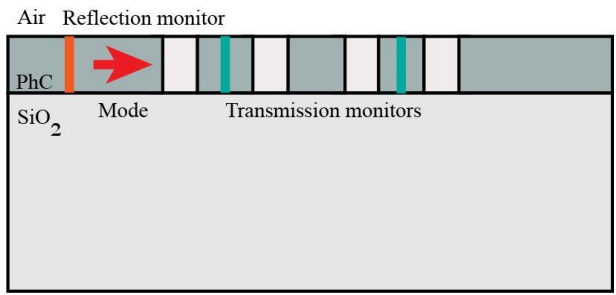 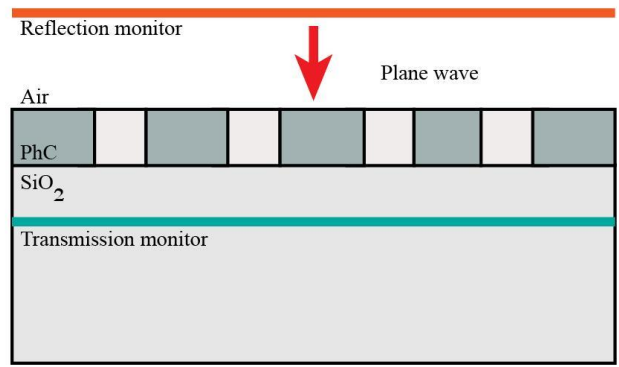

**Fig. S4 Out-of-plane and in-plane schematics.** Schematics of FDTD (a) out-of-plane (b) in-plane simulations.

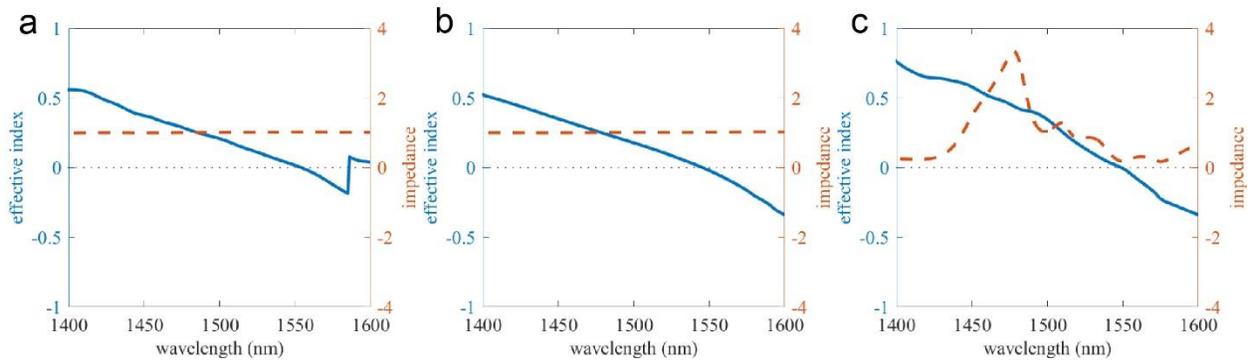

**Fig. S5 Effective index and impedance.** Effective refractive index (blue) and impedance retrieved (orange) from FDTD simulated in-plane complex reflection and transmission coefficients of (a) lossy, (b) resonance-trapped, and (c) symmetry-protected ZIM PhC designs.

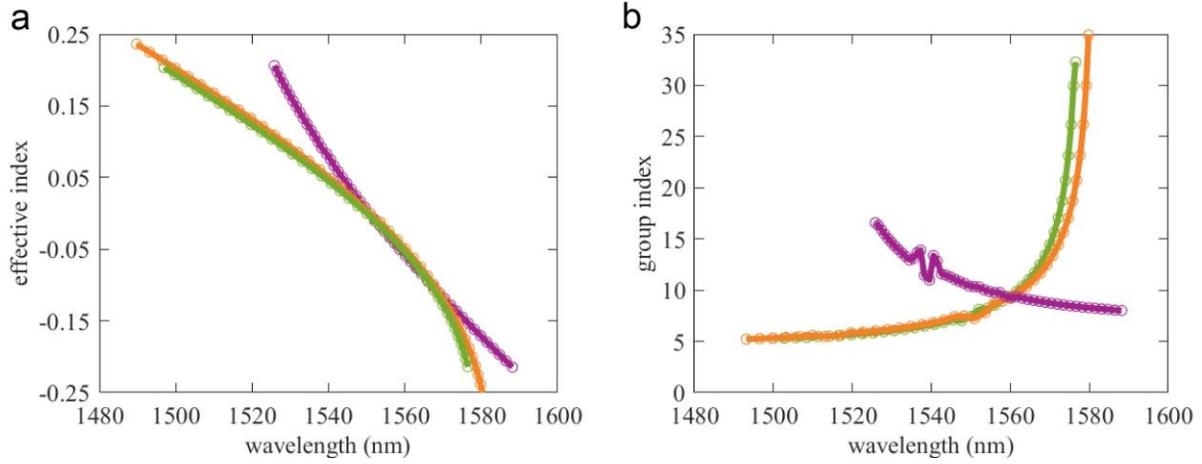

**Fig. S6 Effective index and group index.** (a) Effective index and (b) group index computed from the slope of the Dirac-cone dispersion in Fig. 1 (a, d, g). Both effective index and group index are in the Γ- $X$ direction for lossy (green) and resonance-trapped ZIM PhC (red), and the Γ- $M$ direction for symmetry-protected ZIM PhC (blue).

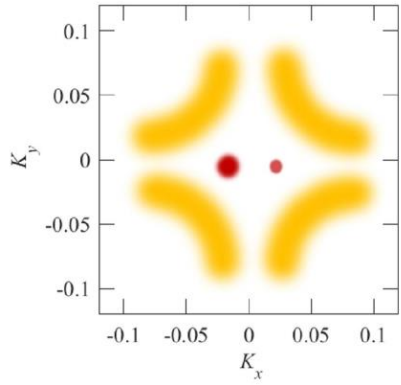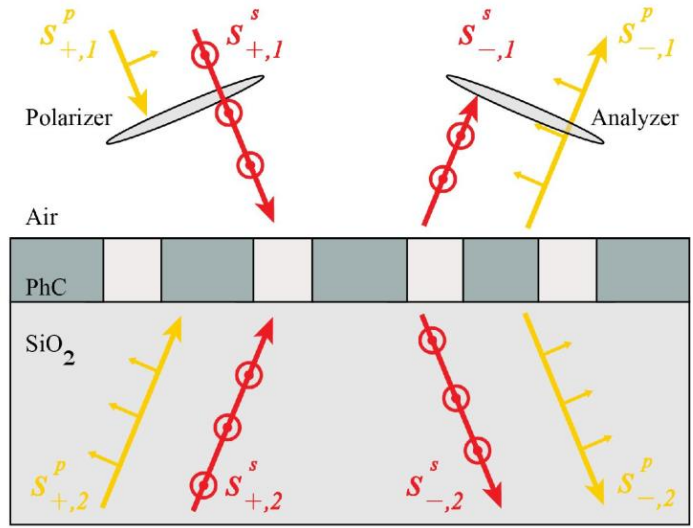

**Fig. S7 Measurement method.** (a) Depiction of the *k*-space contour of PhC slab with on resonant scattering. Brighter red dot (right) is the laser pump point, the darker red dot (left) is the conjugate pump point. (b) Principle of cross-polarization.

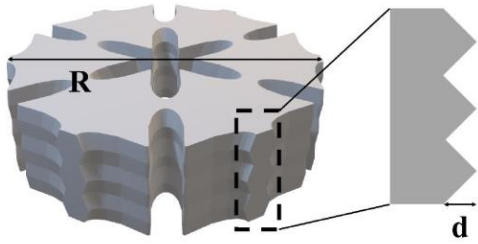 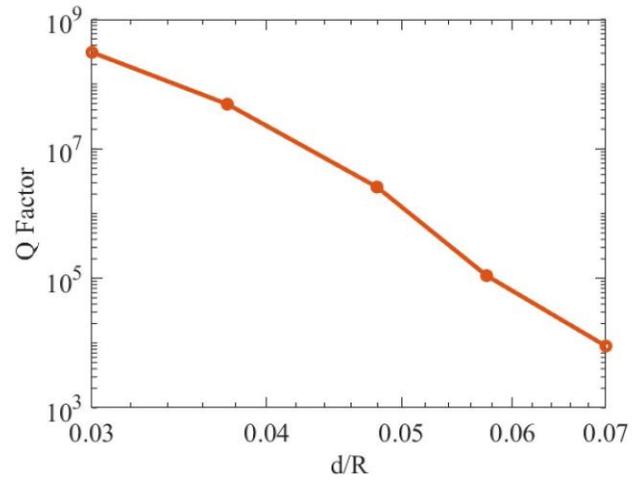

**Fig. S8 Fabrication disorder increases radiative loss**. (a) Sidewall roughness simulated using a triangular saw-tooth pattern of height $d$ to emulate the roughness introduced with reactive-ion etching. (b) COMSOL simulated Q-factor as a function of the fabrication disorder $d/R$ where $R$ is the unit cell pitch.

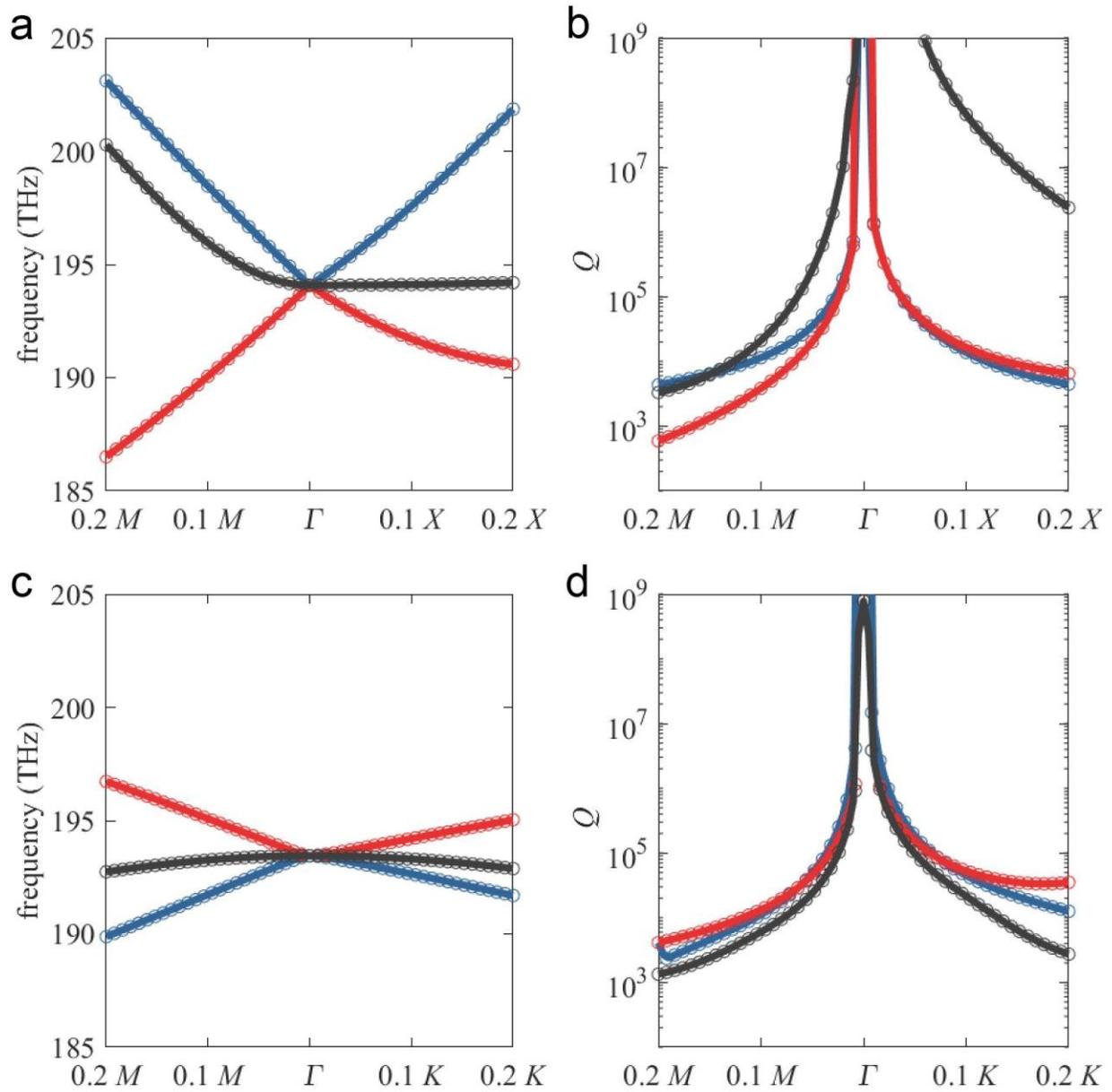

**Fig. S9 Bound states in the continuum (BIC) in suspended devices.** Band structure (left) and $Q$-factor (right) for (a) resonance-trapped and (b) symmetry-protected ZIM PhC designs. The colors in the band structure and $Q$-factor plots correspond to distinct modes. When there is a mirror symmetry in the out-of-plane direction, both resonance-trapped and symmetry-protected ZIM PhCs show diverging $Q$-factors at the Γ point, indicating the formation of a BIC.

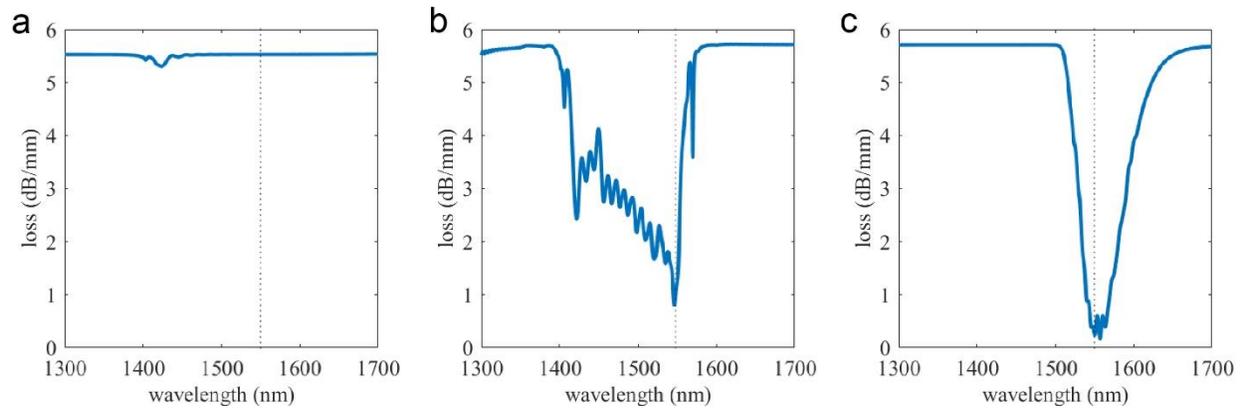

**Fig. S10 Propagation loss.** Lumerical FDTD simulated in-plane propagation losses for (a) lossy, (b) resonance-trapped, and (c) symmetry-protected ZIM PhC slab designs along the $\Gamma$-$X$ (a, b) and $\Gamma$-$M$ (c) directions (Fig. S11(b)). The propagation length of the simulation domain is 10 $\mu m$.

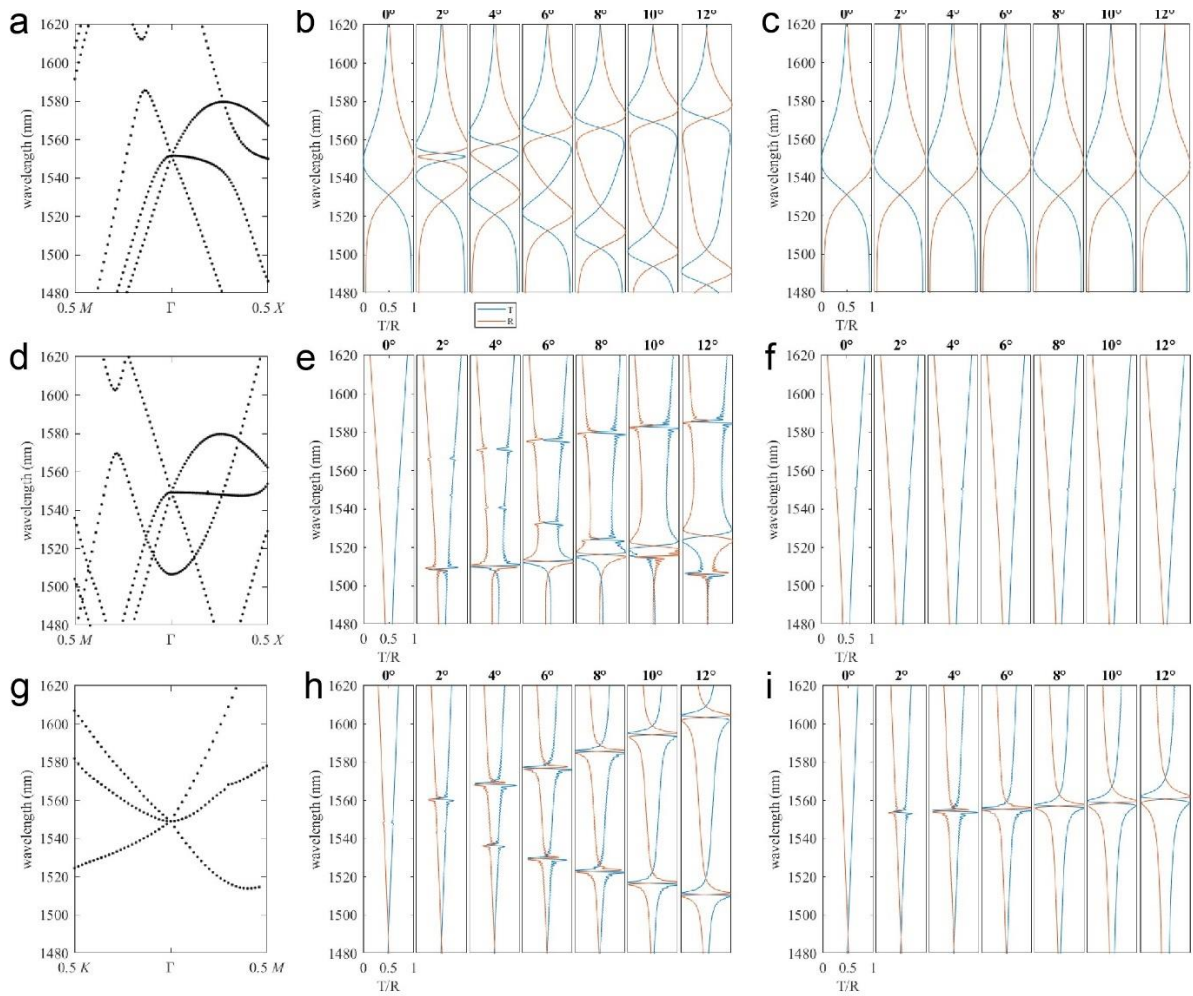

**Fig. S11 Angle-resolved transmission/reflection spectra.** Band structures (left), transmission/reflection spectra at out-of-plane incidence with s-polarized light (middle) and p-polarized light (right) for lossy (top), resonance-trapped (middle) and symmetry-protected (bottom) ZIM PhC designs. As the out-of-plane incidence angle varies, the Fano resonance shifts along with the wavelength. The vanishing of the Fano resonance at the incident angle of zero indicates high-Q modes at Γ point.